\pdfoutput=1
\documentclass[1p]{nmeauth}

\usepackage{amssymb}
\usepackage{bm}
\usepackage{amsmath}
\DeclareMathOperator*{\argmin}{argmin}

\usepackage{subfigure}
\usepackage{graphicx}
\usepackage{color}

\usepackage{siunitx}
\def\GLOBAL_SCALE{0.75}

\usepackage{mathrsfs}

\usepackage[boxed,ruled,linesnumbered]{algorithm2e}

\begin{document}

\title{Projection-based model reduction for contact problems}

\author{Maciej Balajewicz\affil{1}\corrauth, 
        David Amsallem\affil{2}, 
        Charbel Farhat\affil{2}}

\address{\affilnum{1}Department of Aerospace Engineering, 
         University of Illinois at Urbana-Champaign, Urbana, IL, USA\\
         \affilnum{2}Department of Aeronautics and Astronautics, 
        Stanford University, Stanford, CA, USA} 

\corraddr{mbalajew@illinois.edu} 

\begin{abstract} 
To be feasible for computationally intensive applications such as parametric
studies, optimization and control design, large-scale finite element analysis
requires model order reduction. This is particularly true in nonlinear settings
that tend to dramatically increase computational complexity. Although
significant progress has been achieved in the development of computational
approaches for the reduction of nonlinear computational mechanics models,
addressing the issue of contact remains a major hurdle. To this effect, this
paper introduces a projection-based model reduction approach for both static and
dynamic contact problems. It features the application of a non-negative matrix
factorization scheme to the construction of a positive reduced-order basis for
the contact forces, and a greedy sampling algorithm coupled with an error
indicator for achieving robustness with respect to model parameter variations.
The proposed approach is successfully demonstrated for the reduction of several
two-dimensional, simple, but representative contact and self contact
computational models. 
\end{abstract}

\keywords{
contact, greedy sampling method, nonlinear model reduction, non-negative matrix
factorization, reduced-order basis, reduced-order model, singular value
decomposition}

\maketitle

\section{Introduction}
\label{sec:Introduction}

The nonlinear finite element (FE) analysis (FEA) of large-scale systems often
requires prohibitively large computational resources.  These are typically
prescribed by the fine discretization of the computational domain that leads to
a large number of degrees of freedom (dofs) in the system, as well as, in the
case of dynamic analysis, the potentially large number of time-steps needed to
accurately describe the evolution of the system. 

Projection-based model order reduction (MOR) techniques alleviate the first
issue by restricting the solution space to a smaller subspace, thereby reducing
the number of dofs. While many approaches have already been developed for the
efficient reduction of linear computational
models~\cite{moore81,grimme97,willcox01,amsallem11}, three main strategies have
been explored so far for efficiently reducing nonlinear computational models.
The first one is based on linearization techniques~\cite{rewienski06,gu08}.  The
second one is based on the notion of
pre-computations~\cite{barbic_stvenant_05,balajewicz13} but is limited to
polynomial nonlinearities. The third strategy relies on the concept of
hyper-reduction --- that is, the approximation of the reduced operators
underlying a nonlinear reduced-order model (ROM) by a scalable numerical
technique based on a reduced computational
domain~\cite{ryckelynck05,an08,chaturantabut10,carlberg11,amsallem12:localROB,farhat14,amsallem14:smo}.
While several hyper-reduction techniques have been proposed in the literature,
two of them have been designed specifically for the nonlinear FEA of structural
and solid mechanics problems: The \emph{a priori} hyper-reduction
method~\cite{ryckelynck05}, and the Energy Conserving Sampling and Weighting
method~\cite{an08,farhat14,farhat15}. Nevertheless, contact problems remain a
major hurdle for nonlinear model reduction in the context of structural
analysis. This is because, among other things, contact problems are
characterized by inequality constraints that complicate the reduction process. 

Most if not all (nonlinear) computational contact methods proceed in two steps.
The first one focuses on contact detection --- that is, the identification of
nodes, edges, and/or faces of the computational model that should be in contact.
The second step focuses on contact enforcement --- that is, the satisfaction of
the contact constraints defined by physical laws such as non-penetration and
frictional behavior. In practice, the aforementioned constraints are enforced
using one of three popular approaches: The penalty method, the Lagrange
multiplier method, or the augmented Lagrange multiplier
method~\cite{simo92,wriggers06}. Attention is focused in this paper on the case
of the Lagrange multipler method (or its augmented version).

For large-scale nonlinear dynamical systems, proper orthogonal decomposition
(POD)~\cite{sirovich87} is the method of choice for generating the reduced-order
basis (ROB) needed for constructing a ROM. It proceeds by collecting solution
snapshots during a {\it training} procedure, then compressing them using the
singular value decomposition (SVD) method.  The resulting ROB minimizes the
projection error of the snapshots. However, when the Lagrange multiplier method
is chosen for enforcing contact constraints, reducing the contact forces --- or
equivalently, the Lagrange multipliers --- requires special care because the
reduced multipliers must have a positive sign. A ROB for these dual variables
based on SVD would not enforce this positivity requirement \emph{a priori}.  For
this reason, it is proposed in this paper to reduce positive quantities such as
contact forces or Lagrange multipliers using a positive counterpart of the SVD
method known as the non-negative matrix factorization (NNMF) method. Introduced
first in the context of image compression~\cite{lee99}, this method builds
low-rank positive factors that approximate a given non-negative matrix. Here, it
is shown that the proposed usage of NNMF results in the construction of a ROB
that can accurately represent the Lagrange multipliers and lead to the effective
reduction of contact computational models. 

In~\cite{haasdonk12:finance,zhang14}, the authors have addressed similar issues
by constructing a ROB for the Lagrange multipliers using a positive linear
combination of pre-computed snapshots of these dual variables. For
time-dependent problems, this approach can rapidly become impractical as it can
lead to the construction of a ROB of very large dimension. In this work, NNMF
provides a natural procedure for optimally compressing a potentially large
number of snapshots of the dual variables and constructing a small dimensional
ROB for approximating them.

A more generic model reduction issue is the robustness of a ROM with respect to
variations of the model parameters.  Indeed, a ROM is truly useful when it can
be used as a surrogate of the underlying high-dimensional model (HDM) for
parameter values that may be different from those sampled for the purpose of
constructing a ROB. Contact problems can be particularly sensitive to parameter
variations, for example, when the contact areas are very sensitive to such
variations.  A popular approach for constructing a ROM that is valid in a large
region of the model parameter domain is to couple a greedy approach with one or
several~\emph{a posteriori} error
estimators~\cite{veroy05,buithanh08,pdt14,choi15} in order to effectively sample
the parameter domain for computing solution snapshots and constructing a ROB.
Such an approach constructs increasingly accurate ROMs by detecting locations in
the model parameter domain where the errors associated with the ROM are the
largest.  The associated HDM is subsequently reconstructed at the identified
worst-error parameter values and solution snapshots are computed and stored.
Then, the ROB --- and therefore ROM --- is updated based on these additional
snapshots, thereby reducing drastically the error(s) for the newly sampled
parameter values. This procedure terminates whenever the estimated ROM error(s)
is (are) below a specified tolerance throughout the model parameter domain of
interest. It leads to a ROM that remains accurate away from the training
configurations. Therefore in this work, a greedy approach is also developed to
construct both primal and dual ROBs that are robust in a given model parameter
domain. Specifically, the greedy approach developed in this paper relies on the
definition of an error indicator for the contact problem, and successive updates
of both primal and dual ROBs computed using SVD and NNMF, respectively, are
performed.

The remainder of this paper is organized as follows. The notation adopted in
this paper is presented in Section~\ref{sec:notations}. The considered family of
contact problems is derived in Section~\ref{sec:contact}.  The proposed model
reduction procedure for contact problems is described in Section~\ref{sec:MOR}.
This procedure is applied in Section~\ref{sec:applications} to the reduction of
three different contact problems.  Finally, conclusions are offered in
Section~\ref{sec:Conclusions}.

\section{Notation}\label{sec:notations}
Throughout this paper, matrices are denoted by bold capitals (ex. $\bm{A}$),
vectors by bold lower cases (ex. $\bm{a}$), and subscripts identify rows and
columns (ex. $A_{i,j}$ is the entry of $\bm{A}$ located in the $i$-th row and
$j$-th column of this matrix).

$\bm{A}^+$ denotes the Moore-Penrose pseudo-inverse of the matrix $\bm{A}$.

$\bm{I}_{N}$ identifies the identity matrix of size $N$ and $\bm{0}$ identifies
a matrix of zeros.  $\bm{1}_{N}$ identifies the vector of dimension $N$ whose
entries are all ones.

For two matrices $\bm{A}$ and $\bm{B}$ of equal dimension $M \times N$, the
Hadamard product $\bm{A} \odot \bm{B}$ is the matrix of the same dimension whose
entries are given by 

\begin{equation} 
    \left( \bm{A} \odot \bm{B} \right)_{i,j} = A_{i,j}  \cdot B_{i,j} 
\end{equation} 

A discretized variable
$\bm{u}\in\mathbb{R}^N$ at time-step $n \in \mathbb{N}$ is identified by a
superscript as $\bm{u}^{n} \in \mathbb{R}^N$. 

The standard Euclidean norm of a vector $\bm{x}\in\mathbb{R}^N$ and the
Frobenius norm of a matrix $\bm{A}\in\mathbb{R}^{M\times N}$ are denoted by
$\left\| \bm{x} \right\|_2$ and $\left\| \bm{A} \right\|_F$, respectively, and
defined as follows

\begin{equation} 
    \left\| \bm{x} \right\|_2  = \left( {\sum\limits_{i = 1}^N
    {x_i^2 } } \right)^{\tfrac{1} {2}}, 
    \quad \left\| \bm{A} \right\|_F  = \left(
    {\sum\limits_{i = 1}^M {\sum\limits_{j = 1}^N {A_{i,j}^2 } } }
    \right)^{\tfrac{1} {2}}. 
\end{equation} 

Finally,  the negative part of a real number $x$ is defined as $\lbrack x
\rbrack_{-} := \min(x,0)$ and that of a vector $\bm{x}\in\mathbb{R}^N$ is 
defined as $\lbrack \bm{x}\rbrack_{-} = \left[\lbrack x_i \rbrack_{-}\right]$. 

\section{Model contact problem}\label{sec:contact}

\begin{figure}[H]
\centering
\includegraphics{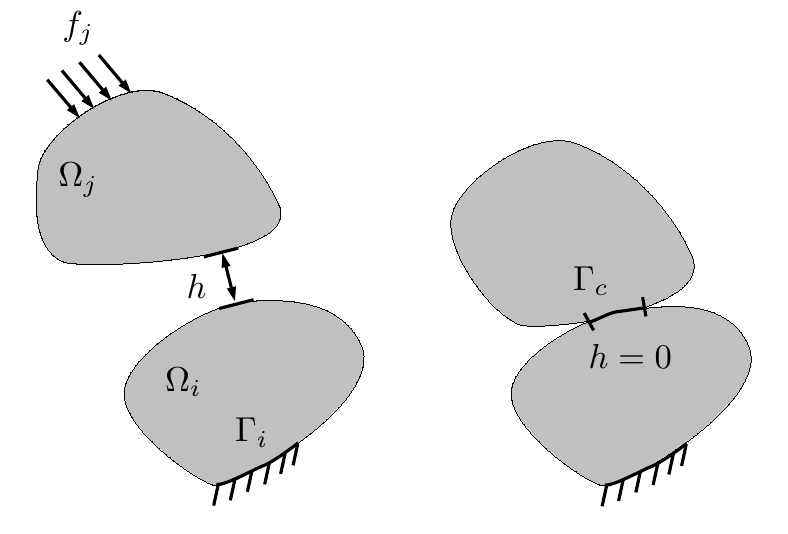}
\caption{Illustration of a generic two-body contact problem.}
\label{fig:generic_contact_problem}
\end{figure}

The main issue addressed in this paper is that of how to reduce the {\it dual}
Lagrange multipliers introduced in a solution process for enforcing the
inequality constraints governing a static or dynamic contact problem. For this
reason, and for the sake of clarity, attention is focused here on a model
contact problem where the individual bodies in contact (for example, see
Fig.~\ref{fig:generic_contact_problem}) have linear material and kinematic
behaviors. The reduction of the {\it primal} displacement solution in the
presence of material nonlinearities and/or large displacements and rotations
raises independent issues that have already been addressed elsewhere in the
literature, for example, in ~\cite
{farhat14,farhat15,amsallem12:localROB,carlberg11}. Furthermore, for simplicity
and without any loss of generality, only the case of a frictionless,
adhesive-free normal contact is considered, and all body problems are assumed to
be undamped and semi-discretized on uniform matching meshes.

Using the modeling assumptions stated above, the FE semi-discretization of a
dynamic $N_{\Omega}$-body contact problem can be written in matrix form as

\begin{eqnarray}
\label{eq:FE_model}
\bm{M} \ddot{\bm{u}} + \bm{K} \bm{u} &=& \bm{f} + \bm{B}^{\rm T} \bm{\lambda}\nonumber\\
		      \bm{B}\bm{u} - \bm{c} &\geq& \bm{0}\nonumber\\
				   \bm{u}(0)&=&\bm{u}_0\nonumber\\ 
			     \dot{\bm{u}}(0)&=&\dot{\bm{u}}_0
\end{eqnarray}
where a dot designates a time derivative, the first semi-discrete equation
expresses the dynamic equilibrium of $N_{\Omega}$ given (flexible) bodies, the
inequality constraint derives from the semi-discretization of the
Hertz-Signorini-Moreau contact conditions that coincide with the
Karush-Kuhn-Tucker (KKT) complementary conditions in the theory of
optimization~\cite{wriggers06}, and the last two equations result from the
semi-discretization of the initial conditions. In Eq. (\ref{eq:FE_model}) above,
$\bm{M} \in \mathbb{R}^{N\times N}$ and $\bm{K} \in \mathbb{R}^{N\times N}$,
where $N$ denotes the total number of primal dofs, are constant symmetric
positive definite or semi-definite block diagonal mass and stiffness matrices,
$\bm{u} = \bm{u}(t) \in \mathbb{R}^{N}$ is a semi-discrete displacement vector
and $t$ denotes time, $\bm{f} = \bm{f} (t) \in \mathbb{R}^N$ is a semi-discrete
force vector, $\bm{B}\in \mathbb{R}^{N_{\lambda} \times N}$, where $N_{\lambda}$
denotes the total number of potentially active contact inequalities, is a signed
boolean matrix which extracts from $\bm{u}$ the pairs of dof governed by a
contact condition, $\bm{c}$ is the vector of initial clearances, and
$\bm{\lambda}$ is the vector of semi-discrete Lagrange multipliers. 

Specifically, $\bm{M}$, $\bm{K}$, $\bm{u}$, $\bm{B}$, $\bm{\lambda}$, and
$\bm{f}$ can be written as

\begin{eqnarray}
	\bm{M} &=&
    \begin{bmatrix}
        \bm{M}_1 & \bm{0}   & \cdots & \bm{0}               \\
        \bm{0}   & \bm{M}_2 & \cdots & \bm{0}               \\
        \vdots   & \vdots   & \ddots & \vdots               \\
        \bm{0}   & \cdots   & \cdots & \bm{M}_{N_{\Omega}}
    \end{bmatrix}, \quad
    \bm{K} =
    \begin{bmatrix}
        \bm{K}_1 & \bm{0}   & \cdots & \bm{0}               \\
        \bm{0}   & \bm{K}_2 & \cdots & \bm{0}               \\
        \vdots   & \vdots   & \ddots & \vdots               \\
        \bm{0}   & \cdots   & \cdots & \bm{K}_{N_{\Omega}}
    \end{bmatrix}, \quad
    \bm{u} = \begin{bmatrix}
	    \bm{u}_1 \\
	    \bm{u}_2 \\
	    \vdots \\
	    \bm{u}_{N_{\Omega}}
    \end{bmatrix} \nonumber\\
    \bm{B} & = & \begin{bmatrix}
  \bm{B}_1 & \bm{B_2} & \cdots & \bm{B}_{N_{\Omega}}
    \end{bmatrix}, \quad
    \bm{\lambda} = \begin{bmatrix}
	    \bm{\lambda}_1 \\
	    \bm{\lambda}_2 \\
	    \vdots \\
	    \bm{\lambda}_{N_{\Omega}}
	    \end{bmatrix}, \quad
    \bm{f} = \begin{bmatrix}
	    \bm{f}_1 \\
	    \bm{f}_2 \\
	    \vdots \\
	    \bm{f}_{N_{\Omega}}
    \end{bmatrix}
\end{eqnarray}
where the subscript $i = 1, 2, \cdots, N_{\Omega}$ designates the body $\Omega_i$.

Again, for simplicity and without any loss of generality, an implicit
time-discretization is assumed in the dynamic case so that solving both static
and dynamic contact problems can be formulated as

\begin{gather}
    \label{eq:dynamic_qp}
    \begin{split}
	    (\bm{u}^n,\bm{\lambda}^n) &= \argmin_{\bm{v} \in {\mathbb R}^N,~\bm{\mu} \in {\mathbb R}^{{+}^{N_{\lambda}}}}
    \frac{1}{2} {\bm{v}}^{\rm T} \bm{A} {\bm{v}} - {\bm{v}}^{\rm T} \bm{b}^n - {\bm{\mu}}^{\rm T} 
    (\bm{B} \bm{v} - \bm{c})
\end{split}
\end{gather}
where $T$ designates the transpose operation, the superscript $n$ designates the
$n$-th time-step $t_n$, $t_0 = 0 < t_1 < \cdots < t_n < \cdots < t_{N_t} =
\mathcal T$ is a discretization of $[0, \mathcal T]$ into $N_t + 1$ time-points, $\bm{A} \in {\mathbb
R}^{N\times N}$ is the block diagonal matrix containing for each body its mass
and stiffness matrices scaled according to the chosen time-integration algorithm
and selected time-stepping strategy, $\bm{b} \in \mathbb{R}^{N}$ contains for
each body the right-hand side vector arising from the implicit
time-discretization of the dynamic semi-discrete equations of equilibrium
governing this body, and $\bm{\lambda}\geq\bm{0}$. In the static case, the
superscript $n$ is dropped, $\bm{A} = \bm{K}$, and $\bm{b} = \bm{f}$.

\section{Model reduction}\label{sec:MOR}
\subsection{Galerkin projection}\label{ssec:Galerkin}

For the model contact problem described above, the standard Galerkin
projection-based MOR method is appropriate. In this method, the primal and dual
components of the solution are approximated in two reduced subspaces represented
here by two pre-computed ROBs $\bm{U} \in \mathbb{R}^{N \times p}$ and
$\bm{U}_{\lambda} \in \mathbb{R}^{N_{\lambda} \times p_{\lambda}}$,
respectively.  This can be written as

\begin{equation}\label{eq:APPROX}
    \bm{u}^n(\bm{\gamma}) \approx \bm{U}\bm{u}_r^n(\bm{\gamma}), \quad \bm{\lambda}^n(\bm{\gamma}) \approx \bm{U}_{\lambda} \bm{\lambda}_r^n(\bm{\gamma})
\end{equation}
where $\bm{u}_r \in \mathbb R^{p}$ and $\bm{\lambda}_r \in \mathbb
R^{p_{\lambda}}$ are the generalized coordinates of the reduced displacement and
Lagrange multiplier solutions, respectively, and $\bm{\gamma} \in \mathcal{D}
\subset \mathbb{R}^{m}$ is a vector of $m$ parameters of the contact problem of
interest. Inserting the above two approximations in the saddle point
problem~\eqref{eq:dynamic_qp} gives

\begin{gather}
    \label{eq:reduced_dynamic_L}
        \begin{split}
		(\bm{u}_r^n,\bm{\lambda}_r^n) &= \argmin_{\bm{v}_r \in \mathbb R^p, ~\bm{\mu}_r \in \mathbb R^{{+}^{p_{\lambda}}}} \frac{1}{2} {\bm{v}_r}^{\rm T} \bm{A}_r {\bm{v}_r} - {\bm{v}_r}^{\rm T} \bm{b}_r -{\bm{\mu}_r}^{\rm T} (\bm{B}_r \bm{v}_r - \bm{c}_r)
\end{split}
\end{gather}
where 

$\bm{A}_r := \bm{U}^{\rm T} \bm{A} \bm{U} \in \mathbb R^{p \times p}$, $\bm{b}_r^n:=\bm{U}^T \bm{b}^n \in \mathbb R^p$, 
$\bm{B}_r:=\bm{U}_{\lambda}^T \bm{B} \bm{U} \in \mathbb R^{p_{\lambda}\times p}$, and $\bm{c}_r: =\bm{U}_{\lambda}^{\rm T} \bm{c} \in \mathbb R^{p_{\lambda}}$.

To ensure the non-penetration condition in the contact ROM, the reduced vector of
Lagrange multipliers must be non-negative --- that is, $\bm{U}_{\lambda}
\bm{\lambda}_r^n \geq \bm{0}$. The approach proposed in
Section~\ref{sec:dual_basis} for satisfying this requirement is to construct a
non-negative ROB $\bm{U}_{\lambda}\geq \bm{0}$. Indeed, since the solution of
the contact problem formulated using the contact ROM delivers a positive vector
of generalized Lagrange multiplier coordinates $\bm{\lambda}_r \ge 0$,
$\bm{U}_{\lambda}\geq \bm{0}$ guarantees in this case that $\bm{U}_{\lambda}
\bm{\lambda}_r \geq \bm{0}$.

Algorithm~\ref{Alg:online} below outlines the Galerkin projection-based MOR
method proposed in this paper for solving the contact
problem~\eqref{eq:dynamic_qp}. To this effect, note that in general, it is
feasible to compute the reduced vector $\bm{b}_r^n$ {\it online}. Specifically,
this can be achieved by pre-computing some relevant small-size quantities {\it
offline}.  For example, consider the case where the prescribed, time-dependent
force vector can be decomposed as $\bm{f}(t) = \bm{L} g(t)$, where $\bm{L} \in
\mathbb R^N$ describes the time-invariant spatial distribution of $\bm{f}$ and
$g(t) \in \mathbb R$ describes its temporal evolution. If time-discretization is
performed using the midpoint rule, $\bm{b}_r^n = \bm{M}_r\bm{u}_r^n + (\Delta
t/2) \bm{M}_r\dot{\bm{u}}_r^n + (\Delta t^2/4) \bm{U}^T\bm{L}g(t_{n+1/2})$,
where $\Delta t$ is the computational time-step. In this case, $\bm{b}_r^n$ can
be efficiently computed at each time-step by pre-computing once for all the
quantity $(\Delta t^2/4) \bm{U}^T\bm{L}$.

\begin{algorithm}
\caption{Online solution of the contact problem~(\ref{eq:FE_model}) using a
    Galerkin projection-based contact ROM \label{Alg:online}}
\SetKwInOut{Input}{input}
\SetKwInOut{Output}{output}
\Input
    {Reduced quantities $\bm{A}_r$, $\bm{B}_r$, $\bm{c}_r$, $\bm{u}_r^0$, and
        $\dot{\bm{u}}_r^0$ 
    } 
\Output
    {Generalized coordinates $\{\bm{u}_r^n\}_{n=1}^{N_t}$, $\{\bm{\lambda}_r^n\}_{n=1}^{N_t}$
        
    }
\BlankLine
\For{$n=1,2,\ldots,N_{t}$}
{    
    Construct the reduced vector $\bm{b}_r^n$\;
    
    Solve the reduced saddle point problem~(\ref{eq:reduced_dynamic_L})
    \begin{equation*}
        (\bm{u}_r^n,\bm{\lambda}_r^n) = \argmin_{\bm{v}_r \in \mathbb R^p,~\bm{\mu}_r \in \mathbb R^{{+}^{p_{\lambda}}}} \frac{1}{2} {\bm{v}_r}^{\rm T} \bm{A}_r {\bm{v}_r} - {\bm{v}_r}^{\rm T} \bm{b}_r -{\bm{\mu}_r}^{\rm T} (\bm{B}_r \bm{v}_r - \bm{c}_r)
    \end{equation*} \;         
}
\end{algorithm}

{\underline {\it REMARKS.}} 
\begin{itemize}
\item In general, the primal ROB $\bm{U}$ enjoys an orthogonality property, and the initial conditions are specified for the
high-dimensional fields $\bm{u}^0$ and $\dot{\bm{u}}^0$ which have physical meanings. Hence, these initial conditions can be 
converted into initial conditions for the generalized coordinates $\bm{u}_r^0$ and $\dot{\bm{u}}_r^0$ using (\ref{eq:APPROX})
and the orthogonality property of $\bm{U}$.
\item Both dual unknowns $\bm{\lambda}$ and $\bm{\lambda}_r$ are auxiliary variables. They do not necessarily need special
	initializations.
\end{itemize}

\subsection{Construction of an optimal primal reduced-order basis}\label{ssec:primalROB}

For the model contact problem described in Section \ref{sec:contact}, the
adoption of {\it global} ROBs for both primal and displacement components of the
solution is appropriate. More complex contact problems featuring material and/or
geometric nonlinearities within the solid bodies call however for adaptive
primal and dual ROBs. Such ROBs can be constructed, for example, using the
concept of {\it locality} introduced in \cite{amsallem12:localROB} which does
not necessarily refer to space or time, but to the region of the manifold where
the nonlinear solution lies. 

In either case, a primal (dual) ROB can be constructed from the compression of
primal (dual) solution snapshots --- that is, primal (dual) components of
solutions of problem~\eqref{eq:dynamic_qp} for different time-instances $t_j$
and different instances $\bm{\gamma}_s$ of the parameter vector $\bm{\gamma}$.
Specifically, for each sampled parameter vector $\bm{\gamma}_s$, $s = 1, \cdots,
N_s$, the computed primal and dual snapshots are gathered in matrices
$\bm{X}^{s}$ and $\bm{X}_{\lambda}^{s}$, respectively, with
$X_{i,j}^s:=u_i^j(\bm{\gamma}_s)$ and
${X_{\lambda}^s}_{i,j}:=\lambda_i^j(\bm{\gamma}_s)$, $j=0,\cdots,N_t$.

In general, the primal ROB is not subject to any particular constraint.
Therefore, it can be constructed by POD~\cite{sirovich87} via the SVD
decomposition of the global primal snapshot matrix
$\bm{X}:=[\bm{X}^1,\ldots,\bm{X}^{N_s}]$.  This corresponds to solving the
optimization problem 

\begin{equation}\label{eq:SVDpb}
\underset{\bm{U}\in\mathbb{R}^{N \times p},\,\bm{V}\in\mathbb{R}^{p \times N_sN_t}}
{\text{minimize}}
\displaystyle \| \bm{X} - \bm{U}\bm{V} \|_F^2
\end{equation}
to compute the low-rank approximation of $\bm{X}$
\begin{equation}
    \bm{X} \approx \bm{U} \bm{V}
\end{equation}

Hence, the ROB $\bm{U}$ is constituted of the first $p$ left singular vectors of the
snapshot matrix $\bm{X}$ and $\bm{V} = \bm{\Sigma}\bm{W}^T$, where $\bm{\Sigma}$
is the diagonal matrix of the first $p$ singular values of $\bm{X}$, and $\bm{W}$ is
the matrix of its first $p$ right singular vectors.

\subsection{Construction of an optimal dual reduced-order basis}\label{ssec:dualROB}
\label{sec:dual_basis}

As emphasized in Section~\ref{ssec:Galerkin}, it is essential to preserve the
positivity of the contact constraints after reduction. For this purpose, it is proposed
here to construct a positive dual ROB $\bm{U}_{\lambda}$ using NNMF~\cite{lee99}. This
corresponding to solving  the optimization problem

\begin{equation}\label{eq:NNMFpb}
\begin{aligned}
	& \underset{\bm{U}_{\lambda}\in\mathbb{R}^{N_{\lambda} \times p_{\lambda}},\, 
                \bm{V}_{\lambda}\in\mathbb{R}^{p_{\lambda} \times N_s N_t}}
              {\text{minimize}}
              & & \| \bm{X}_{\lambda} - \bm{U}_{\lambda}\bm{V}_{\lambda} \|_F^2 \\
   & \quad \quad \text{subject to}
   & & \bm{U}_{\lambda} \geq \bm{0}\\
   &&& \,\, \bm{V}_{\lambda} \geq \bm{0}
   \end{aligned}
\end{equation}
where $\bm{X}_{\lambda}:=[\bm{X}_{\lambda}^1,\ldots,\bm{X}_{\lambda}^{N_s}]$ is
the global dual snapshot matrix. The NNMF algorithm leads to the low-rank approximation of
the dual global snapshot matrix by two positive factors

\begin{equation}
    \bm{X}_{\lambda} \approx \bm{U}_{\lambda} \bm{V}_{\lambda}
\end{equation}

Unlike problem~(\ref{eq:SVDpb}), problem~(\ref{eq:NNMFpb}) does not have a closed form
solution. Consequently, this problem is usually solved using an iterative method that typically
converges to a local minimum. Examples of such methods are the original multiplicative updating rule~\cite{lee99}, 
the alternating non-negativity least-squares method~\cite{kim08_2}, and block coordinate descent algorithms~\cite{kim13}. 

\subsection{Snapshot selection}\label{ssec:snapsSelect}

When many snaphsots are collected for the purpose of constructing primal and dual ROBs with a potential for accurate
approximations of the displacement and Lagrange multiplier fields, respectively, the SVD and NNMF of the global snapshot
matrices $\bm{X}$ and $\bm{X}_\lambda$ become computationally intensive. When $N_s$ different instances of
the parameter vector $\bm{\gamma}$ are sampled, both aforementioned matrices have $N_{s}\times (N_t+1)$ columns. 
In this case, the following remarks are noteworthy:
\begin{enumerate}
\item It is not necessary to store the dual snapshot solutions at those time-steps where there is no contact, as 
these snapshots are zero. Hence, the dimension of the matrix $\bm{X}_\lambda$ can be reduced to the number of time-steps 
at which contact is established, without modifying the accuracy of the resulting ROM.
\item The dimensions of the global snapshot matrices can be further reduced by down-sampling the HDM in time. 
This may be necessary when a very large number of time-steps $N_t$ is computed, for example, when the solution snapshots
are obtained using explicit time-stepping in a relatively large time-window $[0, {\mathcal T}]$.
The temporal down-sampling of the solution snapshots may affect however the accuracy of the resulting ROM as it implies
fewer information for training this ROM.  
\end{enumerate}  

In the remainder of this paper, temporally down-sampled sets of primal and dual
solution snapshots are denoted by $\{\bm{u}^n\}_{n\in\mathcal{S}}$ and
$\{\bm{\lambda}^n\}_{n\in\mathcal{S}_\lambda}$, respectively, where
$\mathcal{S}\subset\{0,\cdots,N_t\}$ and
$\mathcal{S}_\lambda\subset\{0,\cdots,N_t\}$. In Section~\ref{sec:DSC}, a
preliminary study of the effect of temporal down-sampling of the solution
snapshots on ROM accuracy is performed. 

\subsection{Construction of parametrically robust ROBs}
\label{sec:greedy_alg}

In summary, primal and dual ROBs can be constructed by compressing primal and
dual components of solution snapshots computed for some parameter instances
$\bm{\gamma}_s\in\mathcal{D},~s=1,\cdots,N_s$. To this effect, it is first noted
that an~\emph{a priori} sampling of the parameter space may miss certain regions
of $\mathcal {D}$ where the ROM will be inaccurate.  This underscores the
importance of sampling $\bm{\gamma}$ at specific instances $\bm{\gamma}_s$ that
enable the construction of a parametrically robust ROM --- that is, a ROM that
is accurate in the entire parameter space $\mathcal{D}$. 

Finding the best samples in $\mathcal D$ is however a combinatorial problem
whose solution is often intractable.  For this reason, economical greedy
strategies have been developed for this purpose~\cite{veroy05,buithanh08,pdt14}.
Such strategies proceed iteratively by identifiying the parameter samples for
which the error associated with the current ROM is the largest, then sampling
the HDM at these parameter instances, and finally updating the ROM using the
additional HDM solution snapshots. 

Finding parameter instances $\bm{\gamma}_s$ that maximize the error of the
current ROM can be performed by solving directly an error maximization problem
using a gradient-based optimization algorithm~\cite{buithanh08}, or a global
optimization approach and a surrogate model~\cite{pdt14}. Alternatively, a basic greedy procedure~\cite{veroy05}
can be designed for this purpose as follows. Given an {\it a priori 
} set of $N_c$ {\it candidate} 
parameter instances $\left\{\bm{\gamma}^{(1)},\cdots,\bm{\gamma}^{(N_c)}\right\}\subset \mathcal{D}$,
where the superscript and pair of parentheses emphasize here the candidate aspect of
a parameter instance $\bm{\gamma}^{(s)}$ and distinguish it from an effectively sampled parameter instance $\bm{\gamma}_s$,
choose the elements of this set which maximize the norm of the error
\begin{equation}
\displaystyle{||| e(\bm{\gamma}) ||| := \left(\sum\limits_{n=0}^{N_t} || \bm{U}\bm{u}_r^n (\bm{\gamma})
- \bm{u}^n(\bm{\gamma}) ||_2^2\right)^{\frac{1}{2}}}
\end{equation}  
between the HDM solution $\bm{u}^n(\bm{\gamma})$ and the associated ROM solution
$\bm{U} \bm{u}_r^n(\bm{\gamma})$.  In practice however, the set of HDM solutions
$\{\bm{u}^n(\bm{\gamma})\}_{n=0}^{N_t}$ is unknown. Therefore, the above error
is conveniently replaced with an error indicator (that is preferrably
economical). Here, this indicator is based on the following contact conditions:

\begin{gather}\label{eq:KKT}
\begin{split}
    \bm{B} \bm{u}^n - \bm{c}&\geq \bm{0} \quad \text{(non-penetration)} \\
    (\bm{B}^+(\bm{A}\bm{u}^n-\bm{b}^n)) \odot (\bm{B} \bm{u}^n - \bm{c}) &= \bm{0} \quad
    \text{(complementary slackness)} \\
    \bm{B}^+(\bm{A}\bm{u}^n - \bm{b}^n) &\geq \bm{0} \quad \text{(contact force positivity)}
\end{split}
\end{gather} 

Specifically, given a set of ROM solutions
$\{\bm{u}_r^n(\bm{\gamma})\}_{n=0}^{N_t}$, the error indicator proposed in this
paper is

\begin{equation}
	\label{eq:EI}
\mathcal{I}(\bm{\gamma},\alpha_1, \alpha_2, \alpha_2) :=\sum\limits_{n\in\mathcal{J}} \left(\alpha_1 \phi(\bm{r}_1^n (\bm{\gamma}))^2 
                            +\alpha_2 ||\bm{r}_2^n (\bm{\gamma})||_2^2
                            +\alpha_3 \phi(\bm{r}_3^n (\bm{\gamma}))^2 \right)
\end{equation}
where
\begin{eqnarray}\label{eqn:r3}
	\bm{r}_1^n(\bm{\gamma}) &=& \bm{B}(\bm{\gamma}) \bm{U} \bm{u}^n_r(\bm{\gamma}) - \bm{c(\bm{\gamma})} \nonumber\\
 \bm{r}_2^n(\bm{\gamma}) &=& \left(\bm{B}^+((\bm{\gamma})(\bm{A}(\bm{\gamma})\bm{U}\bm{u}_r^n(\bm{\gamma})-\bm{b}^n(\bm{\gamma})) \right) 
                                \odot \left( \bm{B}(\bm{\gamma})\bm{U} \bm{u}^n_r(\bm{\gamma}\right)-\bm{c}(\bm{\gamma})) \\
				\bm{r}_3^n(\bm{\gamma}) &=& \bm{B}^+(\bm{\gamma})(\bm{A}(\bm{\gamma})\bm{U u}_r^n(\bm{\gamma}) 
				- \bm{b}^n(\bm{\gamma})) \nonumber
\end{eqnarray}
$\mathcal{J}$ defines a subset of time-steps at which the proposed error indicator is evaluated, and 
$\phi(\bm{v}):= ||\lbrack \bm{v}\rbrack_{-} ||_2$. The coefficients $\alpha_i$, $i = 1, \cdots, 3$ are adjustable
weights that can be used to emphasize the relative importance of each contact condition. 

Because $\mathcal{I}(\bm{\gamma},\alpha_1,\alpha_2,\alpha_3)$ characterizes the violation of the contact
conditions, it is an indicator for the error associated with the ROM solution
$\bm{U}\bm{u}_r(\bm{\gamma})$. The computational complexity of this error
indicator is reasonable in the sense that its evaluation does not require the
solution of any system of equations. It requires only multiplications.  Of
particular interest is the case $\alpha_2 = 0$ for which the evaluation of
$\mathcal{I}(\bm{\gamma},\alpha_1,0,\alpha_3)$ becomes the most economical.  In fact, the
applications discussed in Section \ref{sec:applications} suggest that the case
$\alpha_1 = 1$, $\alpha_2 = \alpha_3 = 0$ leads to a good error indicator. As
for the time-sampling of the snapshots, the case $\mathcal{J}=\{0,\cdots,N_t\}$
leads to the most accurate error indicator. Unfortunately, it generates an
excessive computational burden when $N_t$ is very large. 

In any case, the reader is reminded that parameter sampling is an integral part
of a training procedure that is performed {\it offline}. Therefore, the
computational complexity of the error indicator (\ref{eq:EI}) does not affect
the performance of the ROM computations to be performed {\it online}. 

The training procedure adopted in this work is summarized in Algorithm~\ref{Alg:greedy_alg}. While it relies on the basic greedy
approach outlined above, this procedure can be accelerated using the techniques described in~\cite{pdt14}.

\begin{algorithm}
\caption{Greedy sampling algorithm for dynamic contact problems \label{Alg:greedy_alg}}
\DontPrintSemicolon
\SetKwInOut{Input}{input}
\SetKwInOut{Output}{output}
\Input
    {Initial sampled parameter instance $\bm{\gamma}_1$, set of $N_c$
    candidate parameter instances $\mathcal{C}=\left\{\bm{\gamma}^{(1)},
    \ldots,\bm{\gamma}^{(N_c)}\right\}\subset \mathcal{D}$, maximal number of
    primal and dual basis vectors $p$ and $p_{\lambda}$, respectively, maximum
    number of greedy iterations $N_{greedy} \geq 2$, convergence tolerance 
    $\epsilon < 1$ 
    } 
\Output{Global primal and dual ROBs $\bm{U}$ and $\bm{U}_{\lambda}$,
    respectively, number of sampled parameter instances $N_s$}
\BlankLine
\For{$N_{iter}=1,\ldots,N_{greedy}$}
{    
    Compute HDM snapshots $\{\bm{u}^n(\bm{\gamma}_{N_{iter}})\}_{n\in\mathcal{S}}$
    and $\{\bm{\lambda}^n(\bm{\gamma}_{N_{iter}})\}_{n\in\mathcal{S}_\lambda}$ and
    store them in local snapshots matrices $\bm{X}^{N_{iter}}$ and
    $\bm{X}_{\lambda}^{N_{iter}}$\;
    
    Accumulate global primal and dual snapshot matrices $\bm{X}$ and
    $\bm{X}_{\lambda}$, respectively \[ \bm{X} = \left[ \bm{X}^1,\cdots,
    \bm{X}^{N_{iter}}\right], ~\bm{X}_\lambda = \left[ \bm{X}_{\lambda}^1,\cdots,
    \bm{X}_{\lambda}^{N_{iter}}\right]\]\; 
    
    Construct a primal ROB $\bm{U}$ of dimension $\leq p$ by compressing $\bm{X}$
    using SVD\;
    
    Construct a dual ROB $\bm{U}_{\lambda}$ of dimension $\leq p_{\lambda}$ by
    compressing $\bm{X}_{\lambda}$ using NNMF\;
    
    \If{$N_{iter}=N_{greedy}$}
    {
        $N_{s}=N_{greedy}$\;
        \textbf{terminate the algorithm}\;
    }
    
    \For{$j=1,\ldots,N_{c}$}
    { 
        Compute ROM solutions $\{\bm{u}_r^n\left(\bm{\gamma}^{(j)}\right)\}_{n=0}^{N_t}$ using 
        Algorithm~\ref{Alg:online} and the current ROBs  $\bm{U}$ and
        $\bm{U}_{\lambda}$\;
        
        Compute the \emph{a posteriori} error indicator
        $\mathcal{I}\left(\bm{\gamma}^{(j)}\right)$\;
    }
                
    Find $\bm{\gamma}_{N_{iter}+1}={\operatorname{argmax}}_{\bm{\gamma} \in
    \mathcal{C}} \, \mathcal{I}(\bm{\gamma})$\;
    
    \If{$\mathcal{I}(\bm{\gamma}_{N_{iter}+1}) >\epsilon \mathcal{I}(\bm{\gamma}_{2})$ }
    {
                $N_s = N_{iter}$\;
                \textbf{terminate the algorithm}\;
    }            
}
\end{algorithm}

\section{Applications}\label{sec:applications}
The model reduction approach proposed in this paper for contact problems is
illustrated here with three simple but representative two-dimensional,
parameterized, model problems. The first one is a static problem of the obstacle
type. The two other ones are dynamic contact problems.  In the first
application, the Lagrange multipliers are approximated using a positive linear
combination of the computed snapshots. Data compression is not performed in this
case because the number of snapshots computed during the training procedure is
small. Hence, this first problem is designed to demonstrate in particular the
effectiveness of the proposed greedy algorithm (Algorithm \ref{Alg:greedy_alg})
for sampling the parameter space. The second model problem considered herein is
a dynamic version of the first problem. Its dynamic aspect gives the opportunity
to precompute a large number of solution snapshots.  Hence, it is suitable for
illustrating the effectiveness of the proposed approach for constructing a ROB
for the Lagrange multiplier field. The third considered model problem is a
dynamic contact problem between two parallel Kirchoff plates. It serves the
purpose of demonstrating the applicability of the proposed model reduction
approach to more generic, parameterized, multi-body contact problems. For the
first two problems, the performance of a constructed ROM is assessed in
``predictive mode'' --- that is, for a problem configuration different from that
used for training the ROM. For the third problem, the performance of a ROM is
assessed in ``reproduction'' mode --- that is, for the same problem
configuration as that used for training the ROM.

In all cases, the relative error of the solution delivered by a ROM is defined
in the static case as

\begin{equation}
    \text{relative error (\%)} :=  \frac{|| \bm{U}\bm{u}_r (\bm{\gamma}) - \bm{u}
    (\bm{\gamma}) ||_2^2}{|| \bm{u} (\bm{\gamma}) ||_2^2 } \times 100 
\end{equation} 
where $\bm{u}(\bm{\gamma})$ is the static HDM solution, and in the dynamic case as
\begin{equation}
    \text{relative error (\%)} :=  \frac{\sum\limits_{n=0}^{N_t} || \bm{U}\bm{u}_r^n (\bm{\gamma})
- \bm{u}^n(\bm{\gamma}) ||_2^2}{\sum\limits_{n=0}^{N_t} || \bm{u}^n(\bm{\gamma}) ||_2^2}  \times 100 
\end{equation} 
where $\bm{u}^n(\bm{\gamma})$ is the dynamic HDM solution.

\subsection{Static problem of the obstacle type}
\label{sec:SPOT}

The model problem presented here is that of the computation of the equilibrium
position of a two-dimensional elastic membrane covering the spatial domain
$(x,y)\in\lbrack0,1\rbrack \times \lbrack0,1\rbrack$, constrained by homogeneous
Dirichlet boundary conditions along all its boundaries, subjected to a uniform
load $f=-10$, and facing a parameterized obstacle.  
This model, parametric, static contact problem
can be described by the inequality-constrained Poisson equation
\begin{gather}
    \begin{split}
        \nabla^2 u &= f\\
        u &\geq g(\bm{\gamma})
    \end{split}
\end{gather}  
where
\begin{equation}\label{eq:obstacle_def} 
    \begin{split}
        g(\bm{\gamma}) = -1 &+ 0.4 e^{-200\left((x-\gamma_1)^2 + (y-0.5)^2 \right)} +\gamma_2 e^{-355.56\left((x-0.7)^2 + (y-0.5)^2 \right)}
    \end{split}
\end{equation}
describes the parameterized obstacle.  The range of interest of the two-dimensional
parameter domain $\mathcal{D} = (\gamma_1,\gamma_2)$ is set to
$\lbrack0.3,0.6\rbrack \times \lbrack0.2,0.6\rbrack$.

The elastic membrane is discretized into $200 \times 200$ finite elements, which
generates an HDM with $40,000$ dofs.  For the training procedure, $\mathcal D$
is initially sampled on a $10 \times 10$ uniform tensor grid to generate a set
of $N_c = 100$ candidate parameter instances.  Then, Algorithm
\ref{Alg:greedy_alg} is applied to construct two global primal and dual ROBs
$\bm{U}$ and $\bm{U}_{\lambda}$, respectively, and their associated ROM. Because
at each iteration of the number of computed snapshots is much smaller than $N_c
= 100$, these snapshots are not compressed. Instead, they are directly used to
gradually construct $\bm{U}$ and $\bm{U}_{\lambda}$. In other words, for this
problem, $p$ and $p_{\lambda}$ are evolved in Algorithm \ref{Alg:greedy_alg} as
$p = k = N_{iter}$. Additional ROMs are also built by randomly sampling ({\it
a priori}) $\mathcal D$, for the purpose of comparing their performance to
that of the ROM delivered by Algorithm \ref{Alg:greedy_alg}. Specifically, $20$
instances $(\gamma_1, \gamma_2) \in \mathcal{D}$ are generated using the Latin
Hypercube Sampling (LHS) method, and primal and dual ROBs of dimension
$p=p_{\lambda}=20$ are constructed using the snapshots computed at these sampled
parameter values. Because of the randomness associated with the LHS samples,
this construction process is repeated 50 times.

\begin{figure}
    \centering 
    \subfigure
    {
        \includegraphics{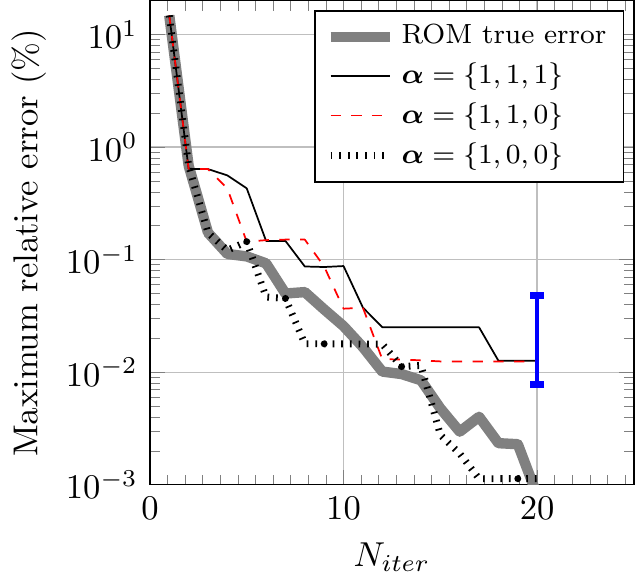}
    }
    \hspace{-0.3cm}
    \subfigure
    {
        \includegraphics{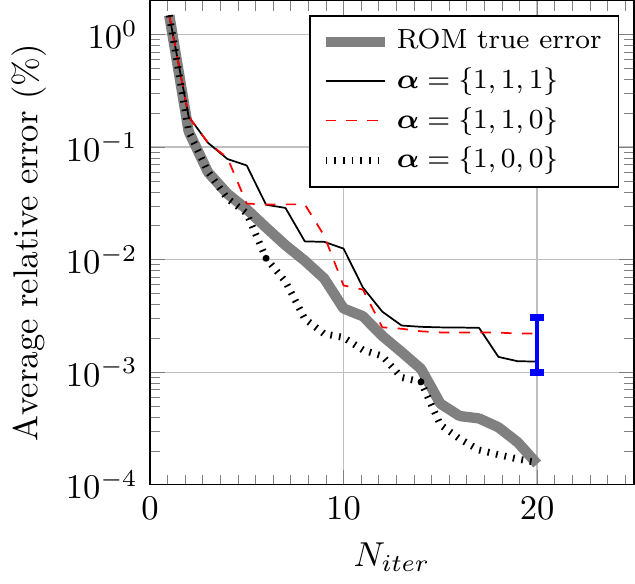}
    }
    \caption{Convergence of Algorithm \ref{Alg:greedy_alg} for the model, 
        parametric, static contact problem of the obstacle type: 
        error bars shown in blue correspond to the ROMs constructed 
        by randomly sampling the parameter
        space and setting $p = p_{\lambda} = 20$.}
    \label{fig:convergence_greedy_steady}
\end{figure}

Figure~\ref{fig:convergence_greedy_steady} reports on the convergence of the
greedy sampling algorithm for this model problem.
The reader can observe that at least in this case, the relatively economical
error indicator $\mathcal{I}(\bm{\gamma},1,0,0)$ performs well and better than
all other considered canonical configurations of this indicator. The reader can
also observe that all intermediate ROMs constructed using the greedy sampling
algorithm outperform the ROMs constructed using the \emph{a priori} random
sampling of $\mathcal D$.  For example, an intermediate ROM constructed with $p
= p_{\lambda} \approx 10$ using Algorithm \ref{Alg:greedy_alg} is shown in
Figure~\ref{fig:convergence_greedy_steady} to deliver the same accuracy as a
twice as large ($p = p_{\lambda}=20$) ROM constructed using random sampling of
the parameter space. Furthermore, the ROM characterized by $p = p_{\lambda} =
20$ and delivered by the same greedy sampling algorithm equipped with
$\bm{\alpha}=\{1,0,0\}$ is found to be an order of magnitude more accurate than
a ROM of equivalent dimension constructed using a random sampling of $\mathcal
D$.

\begin{figure}
\centering
    \subfigure[$\bm{\gamma}=(0.6,0.6)$ 
    \label{fig:static_obstacle_solutions_a}]
    {
        \includegraphics{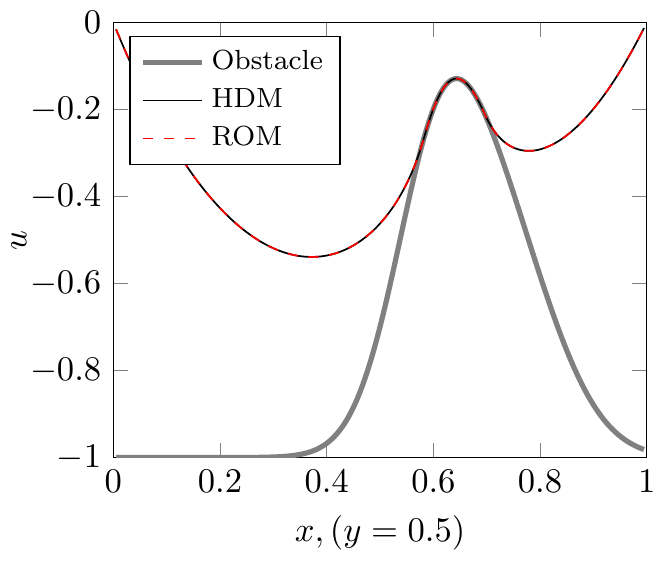}
    }
    \hspace{-0.3cm} 
    \subfigure[$\bm{\gamma}=(0.330,0.377)$ (Maximum ROM error) 
    \label{fig:static_obstacle_solutions_b}]
    {
        \includegraphics{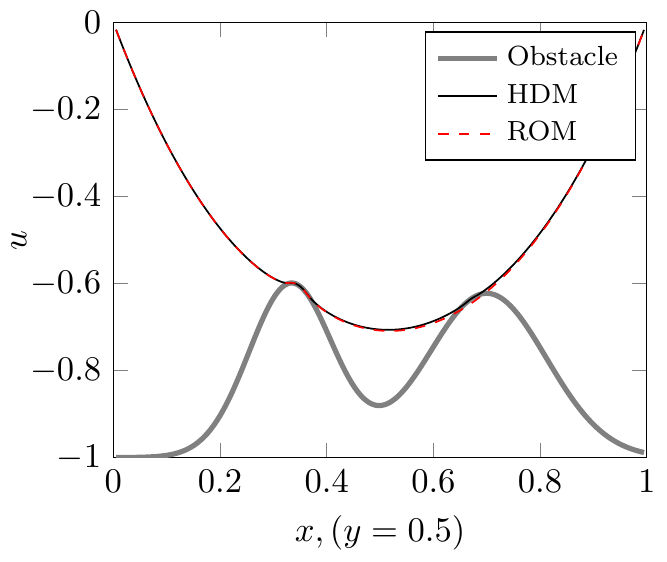}
    }
    \caption{HDM and ROM solutions of the model, parametric, 
            static contact problem of the obstacle type (cut view).}
            \label{fig:static_obstacle_solutions}
\end{figure}

\begin{figure}
    \centering
    \subfigure[HDM\label{fig:XX}]
        {
        \includegraphics[trim = 1.5in 0.5in 13in 0.1in, clip, width=6cm]
        {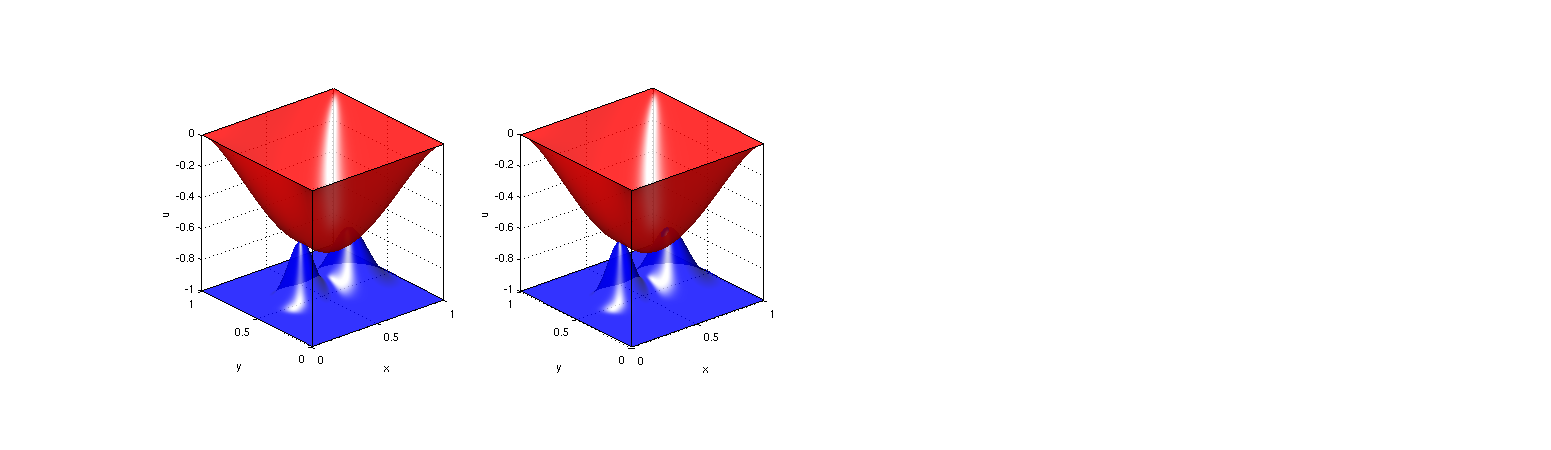}
        }
    \subfigure[ROM\label{fig:XX2}]
        {
        \includegraphics[trim = 5.5in 0.5in 9in 0.1in, clip, width=6cm]
        {2D_steady_greedy_snapshot.png}
        }
	\caption{HDM and ROM solutions for $\bm{\gamma}=(0.330,0.377)$ of the model, 
            parametric, static contact problem of the obstacle type (full view). The obstacle
        and membrane are represented by the blue and red surfaces, respectively.} \label{fig:static3D}
\end{figure}

Figure~\ref{fig:static_obstacle_solutions} showcases the performance of two ROMs
obtained after $20$ iterations of Algorithm \ref{Alg:greedy_alg}. Specifically,
it compares for two different parameter combinations ($\bm{\gamma}=(0.6,0.6)$ in
Figure~\ref{fig:static_obstacle_solutions_a} and $\bm{\gamma}=(0.330,0.377)$ in
Figure~\ref{fig:static_obstacle_solutions_b}) one-dimensional slices of the
two-dimensional HDM and ROM solutions. Note that these two parameter instances
are chosen here for performance assessment for two reasons: (a) neither of them
is part of the set of $N_c$ candidate parameter instances inputted to Algorithm
\ref{Alg:greedy_alg} and therefore neither of them is part of the training of
the constructed ROM, and (b) $\bm{\gamma}=(0.330,0.377)$ is determined {\it a
posteriori} to be the parameter instance for which the constructed ROM has the
largest error. In both cases, the computed HDM and ROM solutions are almost
indistinguishable, thereby demonstrating the accuracy of the proposed approach
for constructing a contact ROM. This accuracy is confirmed in
Figure~\ref{fig:static3D} which compares the full, two-dimensional HDM and ROM
solutions of the considered model, parametric, static contact problem for
$\bm{\gamma}=(0.330,0.377)$.

\begin{figure}
    \centering
    \includegraphics{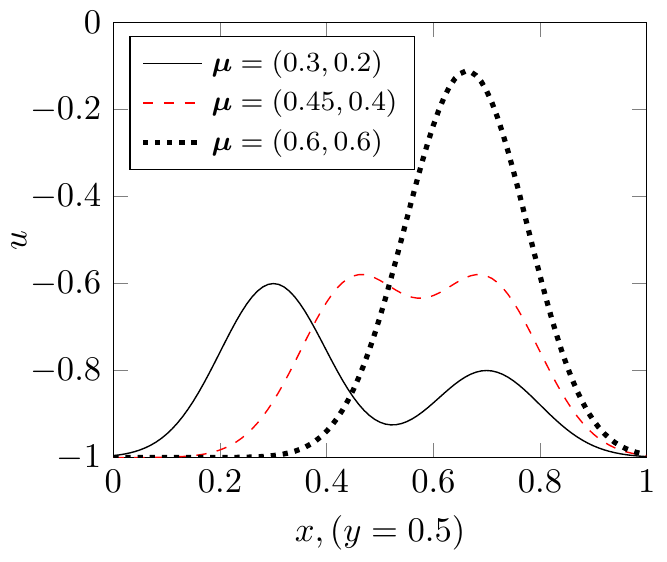} 
    \caption{Parameterized obstacle $g(\bm{\gamma})$.}
    \label{fig:obstacles}
\end{figure}

\subsection{Dynamic contact problem of the obstacle type} 
Next, a dynamic version of the otherwise same model contact problem described in
Section \ref{sec:SPOT} is considered here.  It is described by the
inequality-constrained initial boundary value problem
\begin{gather}
	\label{eq:DOP}
    \begin{split}
        \frac{\partial^2 u}{\partial t^2} &= \nabla^2 u + f\\
        u &\geq g(\bm{\gamma})\\
	    u(x,y:0) &= 0\\
	    \displaystyle{\frac{\partial u}{\partial t}}(x,y:0) &= 0
    \end{split}
\end{gather}  
where $t\in \lbrack 0, 2 \rbrack$. 

The same semi-discrete HDM described in Section \ref{sec:SPOT} is adopted for
this problem. It is discretized in time using the implicit second-order Backward
Differentiation Formula (BDF)
scheme and the constant time-step $\Delta t = 0.005$.  The parameter
domain $\mathcal{D} = (\gamma_1,\gamma_2)$ is set again to
$\lbrack0.3,0.6\rbrack \times \lbrack 0.2,0.6\rbrack$ and initially sampled on
the same $10 \times 10$ uniform tensor grid as before, in order to generate a
set of $N_c = 100$ candidate parameter instances. These are inputted to
Algorithm \ref{Alg:greedy_alg} for performing the training procedure.  At each
$N_{iter}$-th iteration of Algorithm \ref{Alg:greedy_alg}, the solution
snapshots are collected at each time-step. Thus, the number of columns of each
of the global primal and dual snapshot matrices grows in this case as
$(2/0.005)\times N_{iter} = 400 N_{iter}$. The data compression strategy is
chosen so that the size of each of the primal and dual ROBs grows as
$\displaystyle{p = p_{\lambda} =\lfloor 2 + (N_{iter}-1)\frac{98}{9}\rfloor}$.

\begin{figure}
        \centering
        \subfigure
        {
            \includegraphics{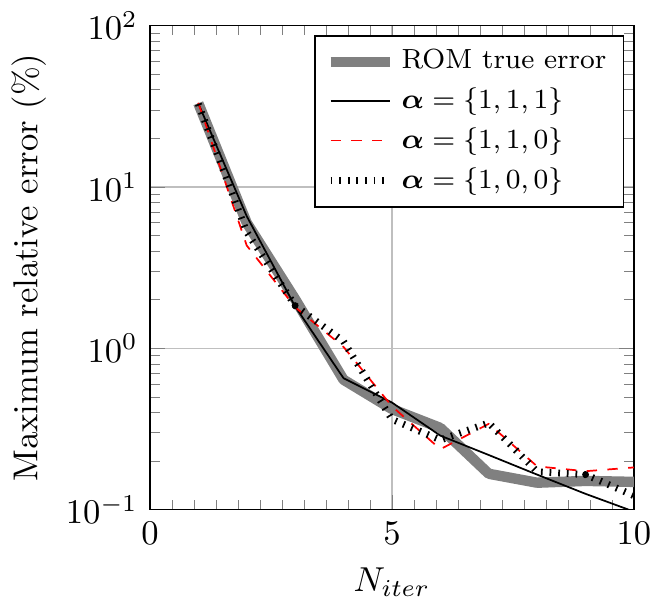}
        }
        \hspace{-0.4cm}
        \subfigure
        {
            \includegraphics{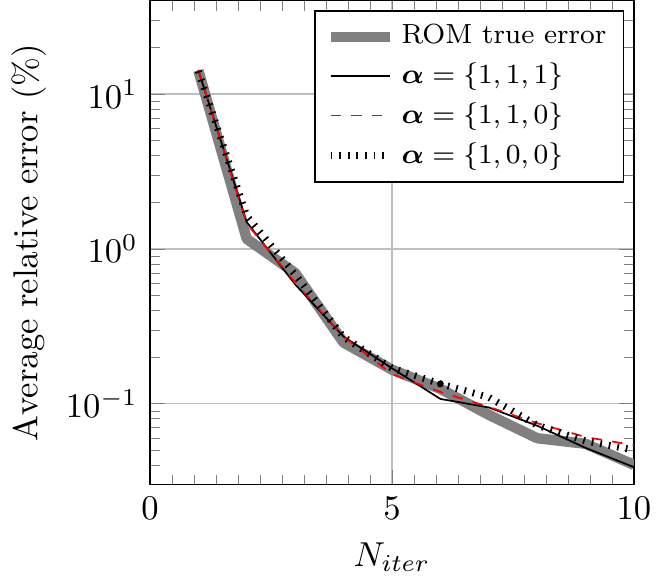}
        }
	\caption{Convergence of Algorithm \ref{Alg:greedy_alg} for the model, 
             parametric, dynamic contact problem of the obstacle type.}
    \label{fig:convergence_greedy_unsteady}
    \end{figure}

Figure~\ref{fig:convergence_greedy_unsteady} reports on the convergence of
Algorithm \ref{Alg:greedy_alg} for this problem. It reveals that in this case,
all considered strategies for $\bm{\alpha}=\{\alpha_1,\alpha_2,\alpha_3\}$
perform equally well. After 10 iterations, each of these strategies leads to a
ROM whose maximum amplitude error is more than 100 times smaller than that of
its counterpart computed at the first iteration.

\begin{figure}
    \centering

    \subfigure[$(x,y) = (0.3,0.5)$.  
            \label{fig:dynamic_obstacle_largest_constact_solution_a} ]
        {
            \includegraphics{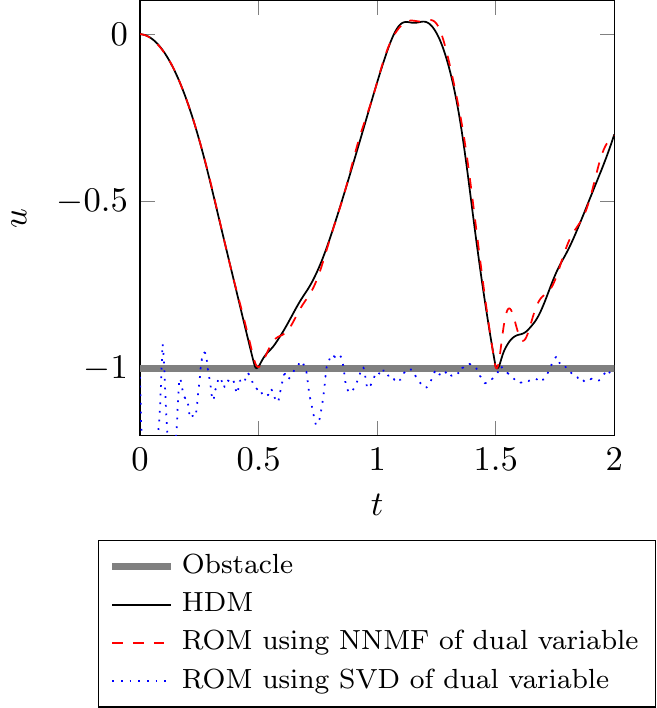}
        } 
        \subfigure[$(x,y)=(0.5,0.5)$  
            \label{fig:dynamic_obstacle_largest_contact_solution_b}]
        {
            \includegraphics{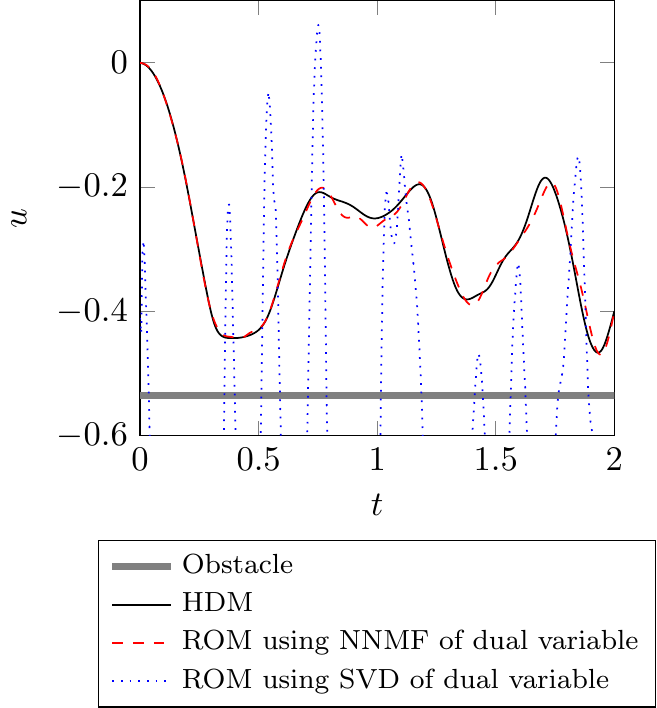}        
        }
	\caption{Time-histories of the HDM and ROM solutions for 
             $\bm{\gamma} = (0.6,0.6)$ of the model, parametric, 
             dynamic contact problem of the obstacle type.}\label{fig:dynamic_obstacle_largest_contact_solution}
\end{figure}

\begin{figure}
    \centering
    \subfigure[$(x,y)=(0.3,0.5)$  
        \label{fig:dynamic_obstacle_max_error_solution_a} ]
        {
            \includegraphics{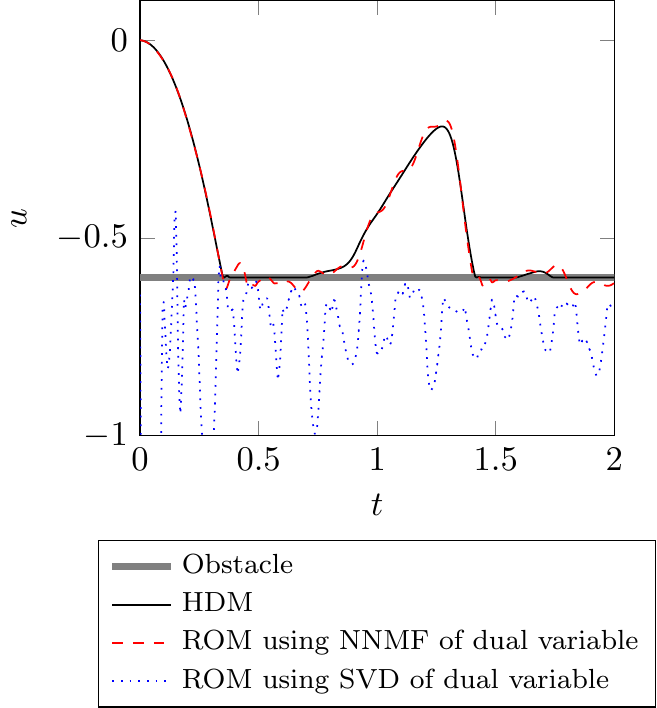} 
        }
    \subfigure[$(x,y)=(0.5,0.5)$
        \label{fig:dynamic_obstacle_max_error_solution_b}]
        {
            \includegraphics{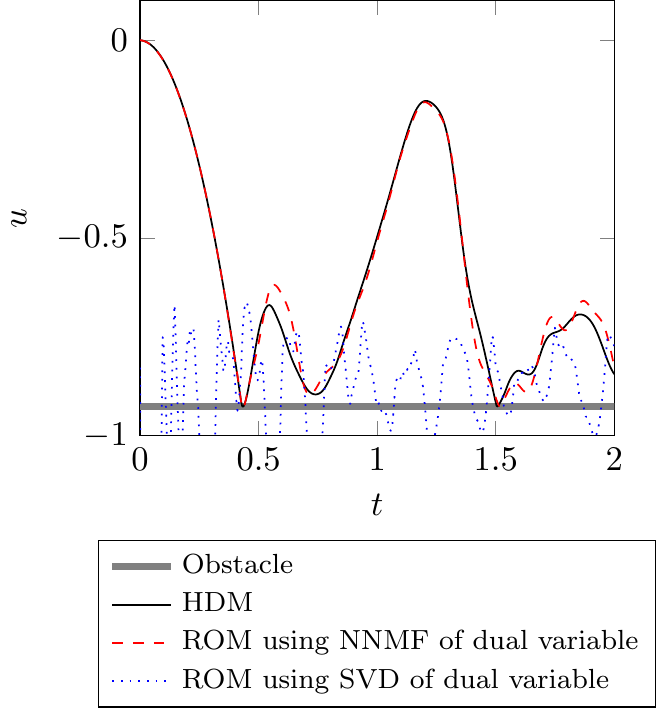} 
        }
	\caption{Time-histories of the HDM and ROM solutions for 
             $\bm{\gamma}=(0.300,0.288)$ of the model, parametric, 
             dynamic contact problem of the obstacle type.} \label{fig:dynamic_obstacle_max_error_solution}
\end{figure}

\begin{figure}
    \centering
    \subfigure[HDM \label{fig:xx} ]
        {
        \includegraphics[trim = 1.5in 0.5in 13in 0.1in, clip, width=6cm]
        {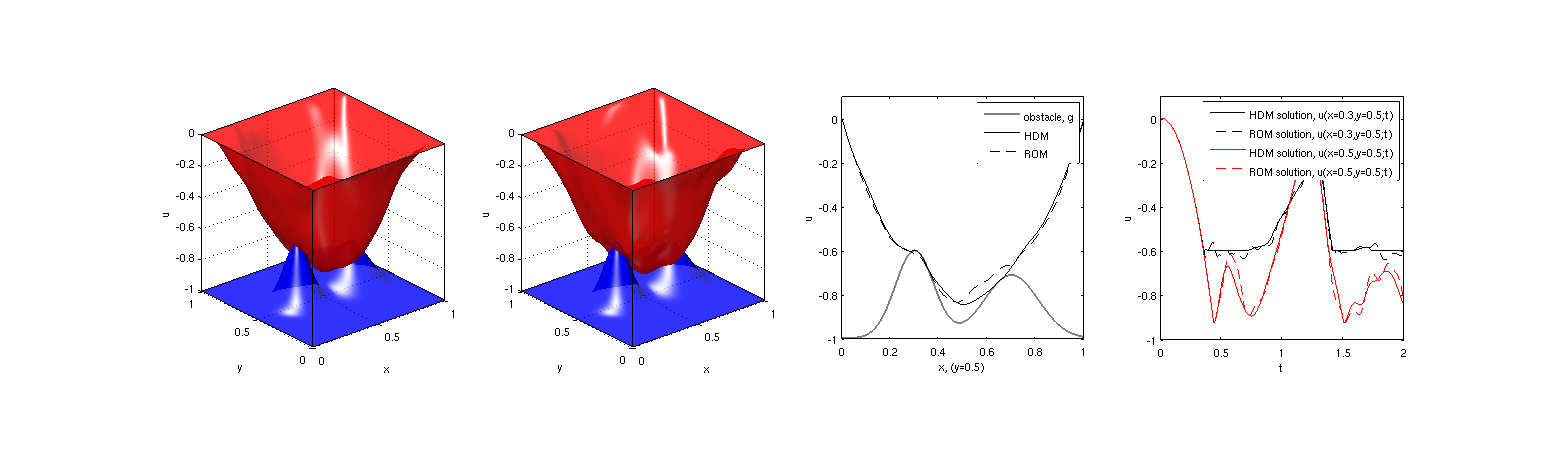}
        }
    \subfigure[ROM\label{fig:yyb}]
        {
        \includegraphics[trim = 5.5in 0.5in 9in 0.1in, clip, width=6cm]
        {2D_unsteady_greedy_snapshot.png}
        }
	\caption{HDM and ROM solutions at $t = 0.2$, $\bm{\gamma}=(0.300,0.288)$, 
             of the model, parametric, dynamic contact problem of the 
             obstacle type (full view). The obstacle
        and membrane are represented by the blue and red surfaces, respectively.}
    \label{fig:dynamic_obstacle_3D}
\end{figure}

Figure~\ref{fig:dynamic_obstacle_largest_contact_solution} and
Figure~\ref{fig:dynamic_obstacle_max_error_solution} display the time-histories
at three different points in space  of the solutions of problem (\ref{eq:DOP})
computed for two different instances of $g(\bm{\gamma})$ using in each case: (a)
the HDM, (b) the ROM obtained after $10$ iterations of Algorithm
\ref{Alg:greedy_alg}, and (c) a variant of this ROM where SVD is used instead of
NNMF to compress the dual snapshots collected within Algorithm
\ref{Alg:greedy_alg}.  Again, note that both parameter instances $\bm{\gamma} =
(0.6,0.6)$ and $\bm{\gamma} = (0.300,0.288)$ are selected here for performance
assessment because neither of them is part of the training of the constructed
ROM, and because $\bm{\gamma}=(0.300,0.288)$ is determined {\it a posteriori} to
be the parameter instance for which the constructed ROM has the largest error.
The reader can observe that as expected, the SVD-based ROMs deliver a poor
performance as they do not satisfy the positivity condition of the Lagrange
multipliers. On the other hand, the NNMF-based ROMs deliver a solid performance.
The solutions they deliver track well the HDM solutions. This solid performance
of the proposed approach for constructing a contact ROM is confirmed in
Figure~\ref{fig:dynamic_obstacle_3D} which focuses on the entire spatial domain.

\subsection{Two-body dynamic contact problem}
\label{sec:DSC}
\begin{figure}
    \centering
    \hspace{3cm}
        \includegraphics{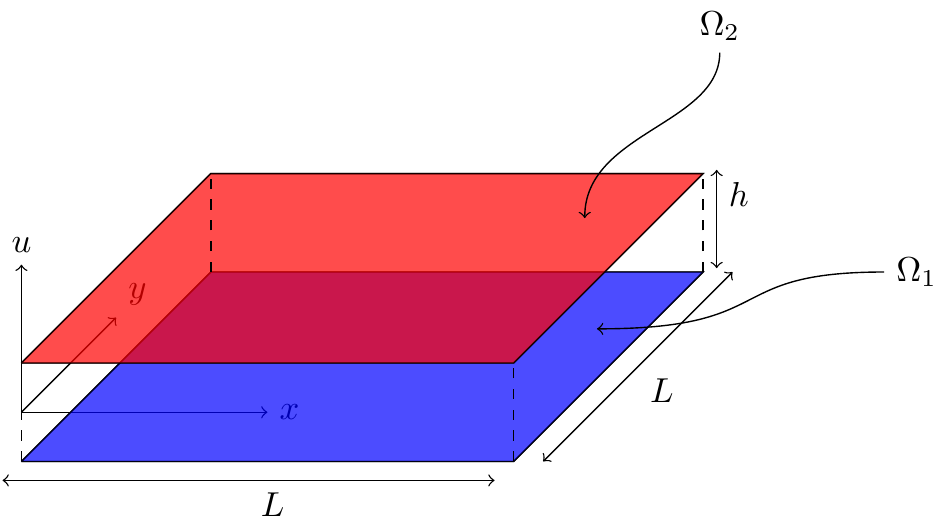}        
	\caption{Two-body dynamic contact problem.}
        \label{fig:dynamic_contact_config}
\end{figure}

Finally, the two-body dynamic contact problem graphically depicted in 
Figure~\ref{fig:dynamic_contact_config} is considered here. 
Each of the two bodies, $\Omega_1$ and $\Omega_2$, is a
homogeneous, isotropic square plate of edge size $L=\SI{1}{\meter}$ and thickness
$h=\SI{1}{\milli\meter}$.  It is modeled as a linearly elastic Kirchhoff-Love plate, and
therefore governed by the partial differential equation 
\begin{equation} 
    \rho h \frac{\partial^2 u}{\partial t^2} + D\nabla^2 \nabla^2 u - f = 0 
\end{equation}
where $\rho$ denotes the density, $u(x,y:t)$ denotes the transverse displacement
field, $f(x,y:t)$ denotes a distributed external force per unit area (pressure), \begin{equation} D = \frac{E h^3}{12(1-\nu^2)}
\end{equation} and $E$ and $\nu$ denote Young's modulus and Poisson's ratio,
respectively. The two plates are assumed to be made of the same material
characterized by $\rho=\SI{7800}{kg/m^3}$, $E = \SI{200}{GPa}$, and $\nu = 0.3$.
One plate is positioned at $h=\SI{4}{cm}$ above the other and is perfectly aligned
with it.  Both plates are clamped and initially at rest.  

The external load per unit area $f$ is applied to the lower plate $\Omega_1$ only, in the upward normal direction. 
It is defined as $f=10^5 f_1(x,y)f_2(t)$, where
\begin{equation}
    f_1 (x,y) = e^{-100\left((x-0.3)^2 + (y-0.4)^2 \right)} 
\end{equation}
\begin{equation} 
    f_2 (t) = H(t-10\Delta t) e^{-1 \times 10^4 (t-10 \Delta t)^2}
\end{equation}
and $H$ denotes the Heaviside function.

Each plate is discretized by $100 \times 100$ finite elements with 1 dof per
node, resulting in a semi-discrete HDM with a total of $20,000$ dofs ($10,000$
dof for each plate). This HDM is
discretized in time using the implicit second-order BDF scheme and the constant
time-step $\Delta t = 2 \times 10^{-4}\SI{}{s}$. 

For this problem, $p$ and $p_{\lambda}$ are set to $p = p_{\lambda} = 40$ and
Algorithm \ref{Alg:greedy_alg} is applied with $N_c = 1$ (no parametric
training) to construct a ROM of size 40. Then, this ROM is applied to the
solution for $t \in \lbrack 0, 0.4\SI{}~{s}\rbrack$ of the same two-body dynamic
contact problem as that solved using the HDM. Note that for $\Delta t = 2 \times
10^{-4}\SI{}{s}$, the time-interval $\lbrack 0, 0.4\SI{}{s}\rbrack$ is sampled
in 2,000 time-steps.

\begin{figure}
    \centering
        \includegraphics{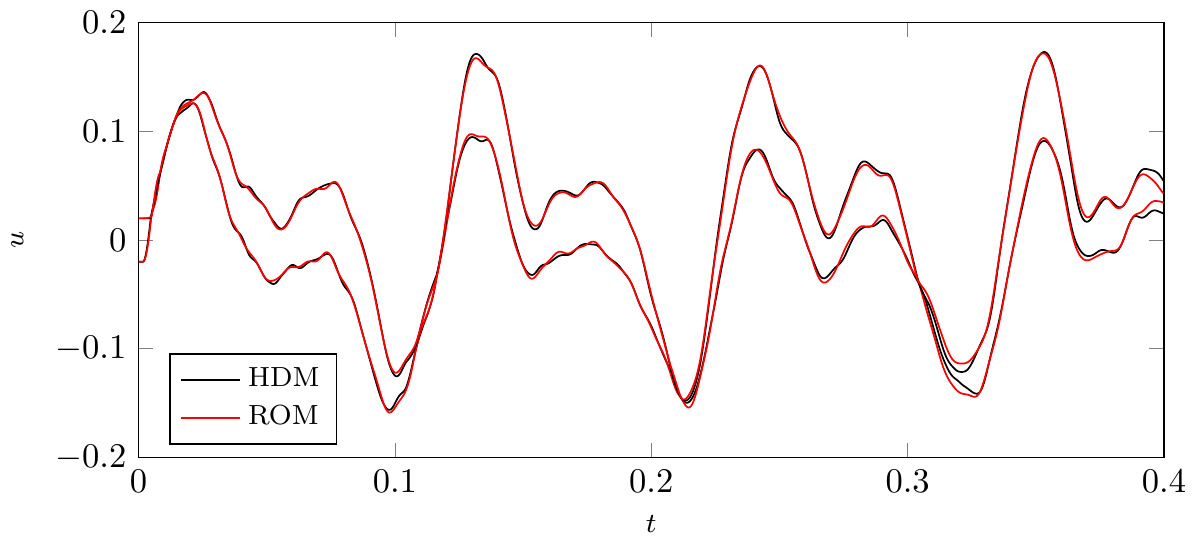}        
	\caption{Time-histories of the HDM and ROM solutions of the 
             two-body dynamic contact problem at $(x, y) = (0.5, 0.5)$.}
        \label{fig:dynamic_contact_solution}
\end{figure}

Figure~\ref{fig:dynamic_contact_solution} displays the time-histories of the HDM
and ROM solutions of this problem at $(x,y) = (0.5,0.5)$.  It shows that the solution
delivered by the constructed ROM tracks remarkably well that computed using the
HDM.

\begin{figure}
    \centering
    \subfigure[HDM]
        {
        \includegraphics[trim = 0.4in 0in 3.95in 0.0in, clip, width=6cm]
        {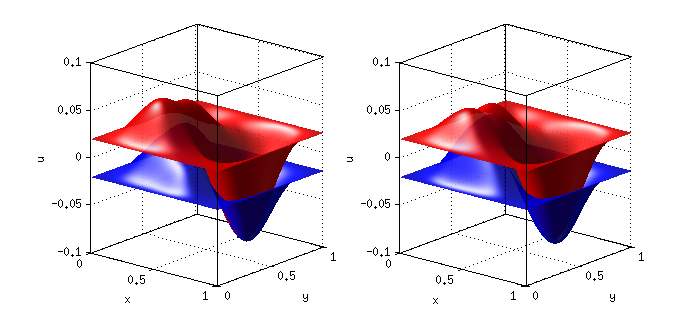}
        }
    \subfigure[$p,k=40$ ROM]
        {
        \includegraphics[trim = 3.85in 0in 0.5in 0.0in, clip, width=6cm]
        {2D_unsteady_plate_snapshot_pk40.png}
        }
    \caption{Snapshots of HDM and ROM solutions of the two-body dynamic 
             contact problem at $t=0.4$.}
    \label{fig:plate_snapshot}
\end{figure}

Figure~\ref{fig:plate_snapshot} complements
Figure~\ref{fig:dynamic_contact_solution} by focusing on the computed HDM and
ROM solutions at a single time-instance, $t=0.4\SI{}{s}$, but the entire
computational domain. It confirms the excellent accuracy of the constructed ROM.

\begin{figure}
    \centering
        \includegraphics{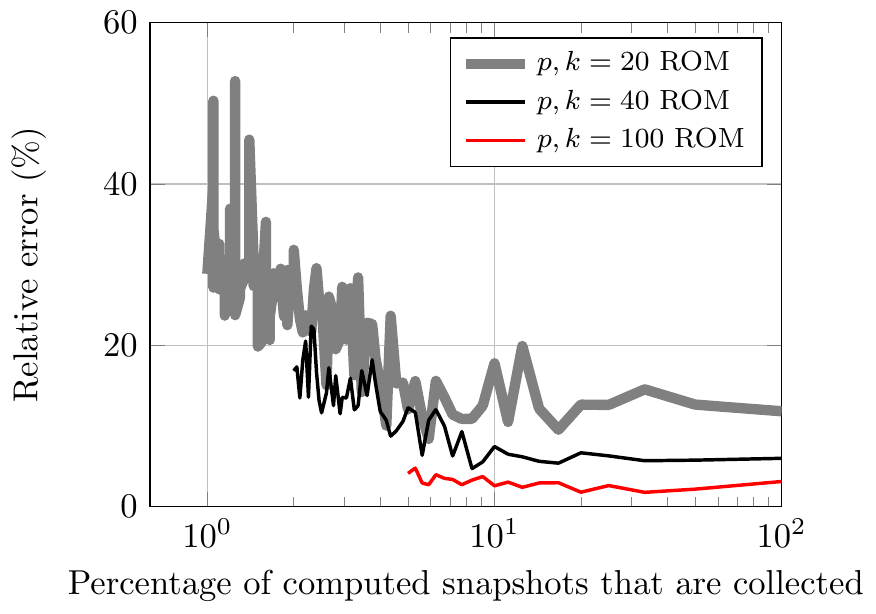}
        \caption{Effect on ROM performance of a down-sampling in time of the 
                 primal snapshots.}
        \label{fig:downsampling_error}
\end{figure}

To illustrate the effect on the performance of a ROM of a down-sampling in time
of the underlying HDM, Algorithm \ref{Alg:greedy_alg} is applied again to the
construction of a series of contact ROMs in which the percentage of computed
solution snapshots that are collected is uniformly decreased. To this effect,
Figure~\ref{fig:downsampling_error} reports the variation of the relative error
of the ROM solution with the percentage of computed solution snapshots that are
collected. The reader can observe that overall, the relative error of the ROM
solution is insensitive to a down-sampling in time, as long as more than 10\% of
the computed solution snapshots are collected for data compression. Beyond this
limit, the relative error of the ROM solution increases sharply to reach the
level of $52.8\%$ when only $25$ equally spaced computed solution snapshots are
collected.

\subsection{Computational speed-up}
All model contact problems discussed above were solved in MATLAB using the
quadratic program solver \verb=quadprog=.  Sparsity was accounted for in all
algebraic entities of all HDMs. In particular, the \verb=interior-point-convex=
algorithm was used to solve all systems of equations arising from all HDMs. On
the other hand, all small-scale and dense systems of equations arising from all
ROMs were solved using the \verb=active-set= algorithm. Numerical experiments
revealed that among all algorithms available in \verb=quadprog=, the
aforementioned solvers are those which offer the best performance for the
considered problems.

All CPU times were measured using the \verb=tic-toc= function on a single
computational thread via the \verb=-singleCompThread= start-up option.  For each
considered contact problem, the speed-up factor delivered by its ROM for the
online computations is reported in Table~\ref{tab:speed_up}.

\begin{table}
    \centering
        \caption{Speed-ups for online computations.}

        \begin{tabular}{ l | c | c | c }
        \hline \hline
            Contact problem        & HDM size   & ROM size    & Speed-up \\
        \hline
	Static, obstacle type      & 40,000 & $p = p_{\lambda} = 20  $  & 865 \\
	Dynamic, obstacle type     & 40,000 & $p = p_{\lambda} = 100 $  & 302 \\
	Dynamic, two-body          & 20,000 & $p = p_{\lambda} = 40  $  & 2,229 \\
        \end{tabular}
        \label{tab:speed_up}
\end{table}

\section{Conclusions}
\label{sec:Conclusions}
The context of this paper is set to that of the model reduction of contact
problems where the contact conditions are enforced using Lagrange multiplier
degrees of freedom (dofs).  In this context, constructing two separate
Reduced-Order Bases (ROBs), one for the primal displacement dofs and one for the
dual Lagrange multiplier dofs, is motivated and justified by the positivity
condition that only the dual variables must satisfy. To this effect, it is shown
in this paper that using the singular value decomposition method and the
non-negative matrix factorization method to compress displacement and Lagrange
multiplier snapshots, respectively, leads to effective primal and dual ROBs and
a promising Galerkin projection method for the reduction of high-dimensional
contact models. For parametric contact problems, it is also shown that the
iterative greedy approach for sampling the parameter domain during the training
of the Reduced-Order Model (ROM) can be equipped with an error indicator of
reasonable offline computational complexity. This error indicator is based on
the residual associated with the non-penetration condition. The computational
complexity of the resulting iterative sampling and ROB construction procedure is
dominated by the cost of a number of high-dimensional simulations equal to the
number of sampling iterations.  Specifically, for two parameterized, static and
dynamic, model contact problems with 40,000 dofs, and one non-parametric
two-body dynamic contact problem with 20,000 dofs, it is shown that the
model reduction approach proposed in this paper and outlined above delivers
online computational speedups in the range of 300 to 2,200. These are promising
results which warrant the extension of the proposed model reduction approach to
more realistic contact problems.

\bibliographystyle{wileyj}
\bibliography{Balajewicz_IJNME_2015}

\end{document}